# Galactic Cosmic Rays and Solar Energetic Particles in Cis-Lunar Space: Need for contextual energetic particle measurements at Earth and supporting distributed observations

Claudio Corti (University of Hawaii, NASA GSFC CCMC), Kathryn Whitman (KBR, NASA JSC SRAG), Ravindra Desai (Imperial College, Warwick), Jamie Rankin (Princeton), Du Toit Strauss (North West University), Nariaki Nitta (LMSAL), Drew Turner (Johns Hopkins APL), Thomas Y Chen (Columbia)

## Synopsis

The particle and radiation environment in cis-lunar space is becoming increasingly important as more and more hardware and human assets occupy various orbits around the Earth and space exploration efforts turn to the Moon and beyond. Since 2020, the total number of satellites in orbit has approximately doubled, highlighting the growing dependence on space-based resources. Through NASA's upcoming Artemis missions, humans will spend more time in cis-lunar space than ever before supported by the expansive infrastructure required for extended missions to the Moon, including a surface habitat, a communications network, and the Lunar Gateway - a space station which will orbit the Moon. Cis-lunar space starts at the top of the Earth's atmosphere and extends out to the orbit of the Moon, including the ionosphere, magnetosphere, free space, and the lunar surface. This paper focuses on galactic cosmic rays (GCRs) and solar energetic particles (SEPs) that create a dynamic and varying radiation environment within these regions. GCRs are particles of hundreds of MeV/nucleon (MeV/n) and above generated in highly energetic astrophysical environments in the Milky Way Galaxy, such as supernovae and pulsars, and beyond. These particles impinge isotropically on the heliosphere and are filtered down to 1 AU, experiencing modulation in energy and intensity on multiple timescales, from hours to decades, due to the solar magnetic cycle and other transient phenomena. SEPs are particles with energies up to thousands of MeV/n that are accelerated in eruptive events on the Sun and flood the inner heliosphere causing sudden and drastic increases in the particle environment on timescales of minutes to days. ***This paper highlights a current and prospective future gap in energetic particle measurements in the hundreds of MeV/n. We recommend key observations near Earth to act as a baseline as well as distributed measurements in the heliosphere, magnetosphere, and lunar surface to improve the scientific understanding of these particle populations and sources.***

# Motivation

## Gaps in Energetic Particle Measurements

The increasing numbers of assets in cis-lunar space, both human and electronic, can potentially experience severe impacts due to the space radiation environment. Previous studies have shown that GCRs in the energy range of 250 MeV/n to 4 GeV/n [1] and SEP protons above 100 MeV [2] are the most significant contributors to radiation dose in humans behind shielding. GCRs and SEPs at these energies and above are able to penetrate the magnetosphere and have implications for the aviation industry. Despite these strong practical motivations, this particle energy range falls in a scientific "gray" area - the low-energy end of the GCR spectrum and the high-energy end for SEPs.

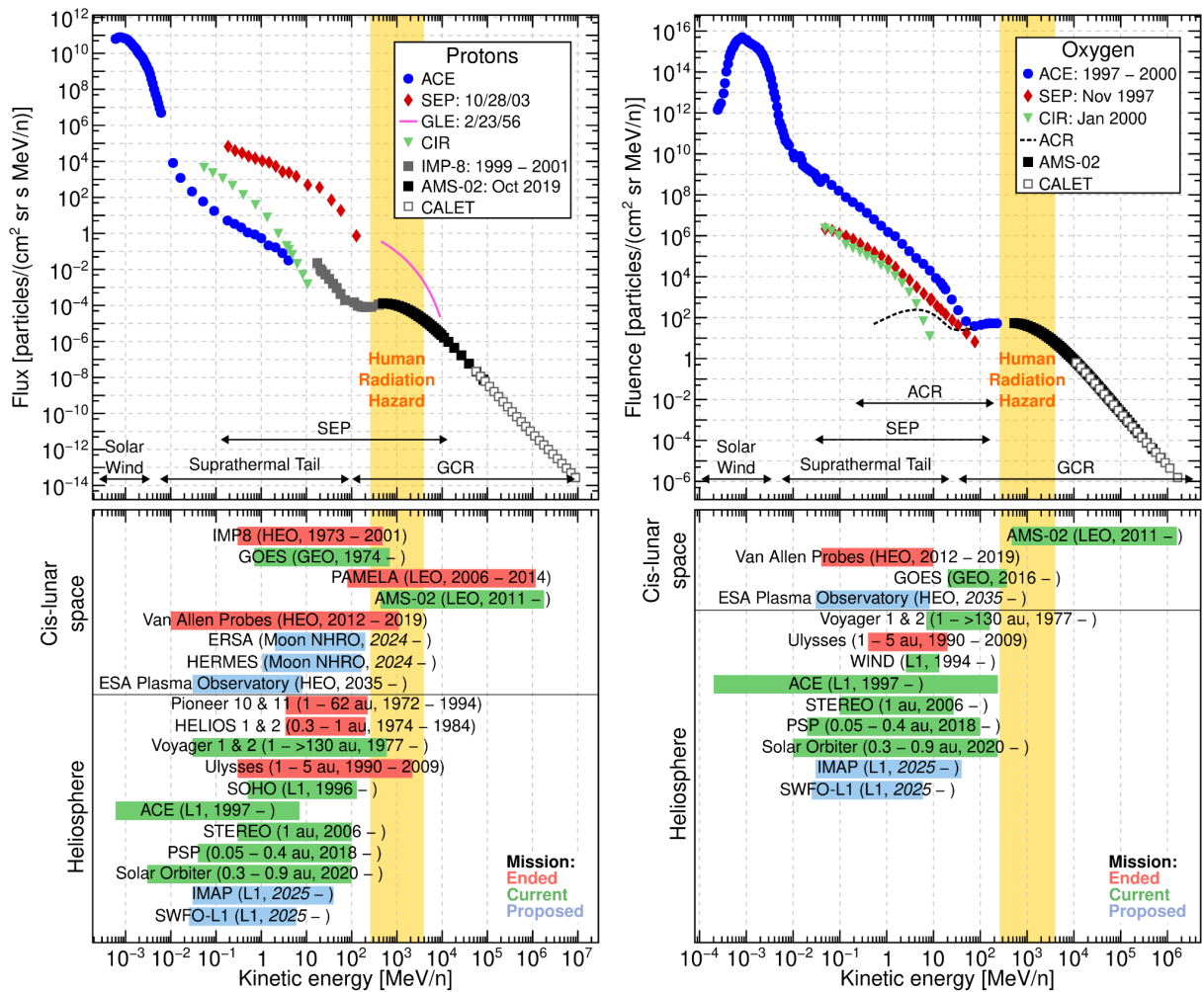

**Figure 1.** Energy ranges of various particle populations (top panels) and *in-situ* particle detectors (bottom panels) in the heliosphere. The vertical yellow band highlights the energy range most relevant for human radiation exposure [1]. The top panels are adapted from [3] and [4]. Other data sources: IMP-8 [5], GLE [6], AMS-02 [7,8], CALET [9,10]. AMS-02 and CALET O fluxes were rescaled to match ACE O fluence for visualization purposes.

Figure 1 shows the energy range of current and proposed in-situ particle detectors, together with the corresponding particle populations in the heliosphere, for protons and oxygen (used as proxy for heavy ions). Only AMS-02 and GOES cover the energy range most relevant for human radiation in cis-lunar space. However, GOES is an operational satellite with limited energy and charge resolution. Cross-calibration with science instruments showed an overestimation of its nominal energy channels [11,12,13,14], high GCR background, and residual particle contamination in low-energy channels [15].

SEP protons with energies above 1 GeV are currently measured only by AMS-02 and indirectly by ground-based neutron monitors (NMs). However, AMS-02 might miss the event's onset due to its orbital location, while not all of the most energetic SEPs lead to a significant signal in NMs. SEP measurements of Z>1 ions are currently limited to the energy range of ACE/SIS, because ACE/CRIS was not designed to operate during high solar activity periods. AMS-02 observed some He SEPs above 250 MeV/n [16], but it is sensitive to heavy ions only above 450 MeV/n. Since no measurement of SEP Z>1 ions exists above 200 MeV/n, this is a largely unexplored area. None of the currently proposed missions by NASA and ESA will have particle detectors covering the range above 200 MeV/n. In particular, it appears that currently-planned future instruments dedicated to space weather (*e.g.*, SWFO-L1) will measure lower energies than current missions. ***In general, no current or proposed mission is capable of continuous measurements of the high-energy end of the SEP spectrum for all elements.***

After ACE is decommissioned, there will be no instrument capable of measuring accurate heavy ion fluxes above 100 MeV/n, creating a gap in the historical set of GCR measurements. This gap is also present in measurements of anomalous cosmic rays (ACRs), which are formed when ionized interstellar neutral matter is accelerated in the outer heliosphere before propagating back to the inner heliosphere [17]. ACE/CRIS data are also used to drive the GCR models from NASA and ESA [18,19].

A gap in energy coverage is also present for GCR electrons and positrons. While current and planned future missions are capable of measuring SEP electrons up to few tens of MeV (*e.g.*, SOHO: 10 MeV [20]; IMAP: 40 MeV [3]), only PAMELA [21,22] (now ended) and AMS-02 [23] provide a continuous measurement of electrons and positrons above 70 MeV and 500 MeV, respectively. Below 100 MeV, measurements made in the heliosphere can be compared directly with Voyager observations in interstellar space [24]. Additionally, when magnetically connected to Jupiter, instruments can detect Jovian electrons between 1 and 100 MeV: these are accelerated in the Jovian magnetosphere and can fill the inner heliosphere [25,26]. No current or future mission will measure electrons and positrons between 40 and 500 MeV.

***A unique opportunity therefore exists for a future mission with an optimized set of measurements to simultaneously return cutting-edge science and fill critical gaps for space weather, as described below.***

## Open Scientific Questions

### 1. GCRs through the Heliosphere, Solar Wind Transients, and Atmosphere

GCRs interact with the turbulent interplanetary magnetic field. Concurrent observations of plasma and GCRs at different locations in the heliosphere spurred the birth of global models of solar modulation, constraining both turbulence and particle transport models [27,28]. However, recent proton, electron, and positron observations suggest that the diffusion and drift coefficients do not always behave as expected from theory [29,30,31,32]. *Continuous measurements of ions and electrons/positrons above 100 MeV/n are needed to improve the understanding of global transport processes in the heliosphere.*

Solar wind transients, like interplanetary coronal mass ejections (ICMEs) and corotating interaction regions, cause a temporary reduction of GCRs called Forbush decreases (FDs) [33,34]. The specific processes at work in these decreases depend on the nature of the transients, and their effects vary with the heliocentric distance [35]. High-cadence measurements of GCR spectra and plasma properties are needed to inform models of turbulence and particle transport inside these propagating structures.

GCR ions are the source of NM counts. Space-borne measurements of the incident GCR flux and composition allow testing and validating NM response functions [36], [37], so that NM data can be used to estimate the dose rate at aviation altitudes [38], as recommended by the International Civil Aviation Organization [39]. It is important to continue cross-calibrating NMs with GCR measurements in space.

### 2. Unusual Trends in GCR and ACR Modulation in Recent Solar Cycles

The past two solar cycles exhibited highly unusual trends with regard to solar modulation when compared to historical solar minima. For example, the 2009 solar minimum produced the highest fluxes of cosmic rays of the space age [40] as well as record-setting rates in NMs [41]. The exact reasons for these high GCR levels are still debated, but these exceptionally low levels of modulation are most likely due to the very quiet and less turbulent solar conditions during recent solar minima, leading to more efficient GCR diffusion [42]. More recently, the 2019 solar minimum was amongst the weakest ever observed, and new records of cosmic ray fluxes were set for the heavy elements ($5 \leq Z \leq 28$) in the mid-energy range (~50 to ~500 MeV/n) [43]. Yet, although current GCR trends reveal strong deviations from their historic values, ACR trends remain historically consistent. The reasons for these deviations and especially for the modern ACR-GCR discrepancy are not currently well understood [44]. Long-term comparisons between ACRs and GCRs will improve understanding of particle transport (*e.g.*, by comparing effects due to different charge states) and acceleration (*e.g.*, by cross-examining particle acceleration in CMEs vs. the termination shock vs. supernova).

*Given the unusual recent trends, it is therefore both critical and timely to continue synoptic measurements of ACRs and GCRs into future solar cycles.*

## 3. Solar Energetic Particle Acceleration and Transport

SEPs are generally classified as either *impulsive* or *gradual*, being associated with acceleration due to either magnetic reconnection in solar flares or diffusive shock acceleration at shocks driven by coronal mass ejections, respectively [45,46,47]. However, some events show characteristics of both processes [48] and the flare *versus* CME origin question remains a topic of investigation. Once accelerated, SEPs propagate through the turbulent interplanetary medium. These processes modify the accelerated SEP spectrum, masking the characteristics of the acceleration processes and making a clear distinction between flare and CME acceleration difficult. Simulations show, for instance, that pitch-angle scattering can modify the SEP spectrum during propagation [49], while observations show several spectral features that might point to the underlying acceleration process [50]. Widespread SEP events [51] are either a result of a very broad acceleration source or indicate that SEPs can propagate efficiently across the mean magnetic field. This may be due to either perpendicular diffusion [52], drift effects [53,54], or a combination of these processes. This acceleration *versus* transport origin of widespread events also remains unsolved. ***New measurements at higher energies, with improved energy and angular resolution, along with direct particle anisotropies*** [55,56]***, will aid in addressing these research questions. Additionally, measurements of relativistic electrons hold value for in-situ forecasting of SEP protons*** [15]***.***

## 4. Magnetosphere and Van Allen Radiation Belts

The Van Allen radiation belts feature protons and electrons with energies up to GeV and ~10 MeV, respectively, and fluxes varying across multiple orders of magnitude. While significant progress has been made in understanding individual processes, how the various contrasting sources and sinks combine to produce the observed radiation variability remains to be explained [57,58].

SEP events can be effectively amplified as incident interplanetary shocks can rapidly inject SEPs into the inner magnetosphere where they remain trapped for extended periods [59]. The ability to model and forecast such extreme events remains an outstanding limitation in radiation belt transport theories and forecasting capabilities. The magnetospheric cavity also reduces in extent by up to a factor of two during geomagnetic storms, which increases SEP access to mid-latitudes [60,61]. This raises radiation levels for near-Earth orbiting infrastructure and astronauts, but also for aviation flights at higher latitudes [62]. As we look forward, missions entering cis-lunar space will have to pass through the Van Allen radiation belts for extended periods due to the increasing use of electric orbit raising maneuvers [63], as for the Lunar Gateway in 2024 [64]. ***Radiation belt forecasts are currently in their infancy and their development requires continuous measurement across a wide range of radial distances.***

# Recommendations

Key Energetic Particle Measurements at 1 AU near Earth

| Key Measurement | Requirements and Possible Implementations |
|---|---|
| Ion energy spectrum from 1 MeV/n to few GeV/n.<br>Time resolution of minutes.<br>Elemental, isotopic, and ionic composition.<br>*See OSQ 1, 2, 3.* | ● Charge and mass resolution better than 1 unit: [65,66].<br>● No saturation: trigger rate of few kHz [8]; or restricted geometry dedicated to high-rate periods [67,68,69].<br>● High-energy range: Cherenkov detector [70]; or tracking spectrometer [8,68,71].<br>● Background suppression: passive and active anti-coincidences [8,68,69,71].<br>● 20% statistical error on 5-minute proton flux at ~1 GeV during FDs and on daily O flux at ~400 MeV/n during solar maximum: effective acceptance >10 cm² sr [8,68,71]. |
| Electron and positron energy spectrum from 1 MeV to 1 GeV.<br>Time resolution of minutes.<br>*See OSQ 1, 3.* | ● Lepton vs hadron discrimination: transition radiation detector, and/or Cherenkov detector, and/or electromagnetic calorimeter [8,70,71,72].<br>● Charge-sign discrimination: tracking spectrometer [8,68,71, 72].<br>● Background suppression: passive and active anti-coincidences [8,68,69,71,72].<br>● 20% (4%) statistical error on daily positron (electron) flux at ~1 GeV during solar maximum: effective acceptance >10 cm² sr [8,68,71, 72]. |
| Ion and electron anisotropy.<br>*See OSQ 3.* | ● Sectored telescope, spinning spacecraft [3,67,73]; or incoming direction measurement [8,68,71, 72]. |
| Solar wind plasma and magnetic field.<br>*See OSQ 1, 2, 3.* | ● Standard solar wind and magnetic field instrument suite. |

**Table 1**. List of recommended key measurements for a long-term baseline reference at 1 AU. OSQ: Open Scientific Questions covered in the previous section.

***In addition to filling the critical energy gap in the cis-lunar environment, a set of dedicated key measurements at 1 AU would continue in the provision of reference measurements that are vital to the larger heliophysics fleet.*** For example, recent findings from PSP suggest that ACR and GCR transport near the Sun behaves much differently than anticipated or explained by current models [74,75], yet these studies required the deconvolution of radial and temporal effects, which could only be achieved by utilizing an appropriate 1 AU baseline

(*e.g.*, ACE, STEREO, and SOHO). In a previous era, Ulysses surprisingly found a significant North-South heliospheric asymmetry in GCR propagation [76]. However, the physical cause and implications related to these observations remains an open question. Measurements by Solar Orbiter as it progresses out of the ecliptic (and any future missions utilizing high-latitude trajectories) will undoubtedly lead to new insights and discoveries, but ***concurrent, reliable measurements from a 1 AU ecliptic baseline will be vital for achieving a full, spatial understanding of transport effects deconvolved from time variations.***

Table 1 lists the set of key energetic particle measurements recommended for a long-term baseline reference at 1 AU, together with their instrumentation requirements and the scientific questions addressed. The main goal of these measurements is to bridge the energy gap between the low-energy end of the GCR spectrum and the high-energy end for SEPs. In addition, capabilities like elemental, isotopic, and ionic composition, time resolution of the orders of minutes, and particle anisotropies will improve our understanding of the open scientific questions described above. It is important that these measurements overlap in time with the current fleet of aging missions (*e.g.*, SOHO, ACE, STEREO, AMS-02), so that a cross-calibration between old and new instruments can be performed. Since energy and species ranges overlap between SEPs and GCRs, the same instrumentation can measure both. In this case, particular attention should be paid to avoid saturation during periods of intense solar activity.

## Distributed Measurements in Cis-Lunar Space

A set of key measurements near Earth, as described above, should be supported by distributed measurements throughout the inner and outer heliosphere and within the magnetosphere to better sample all of the key physics elements and spatial scales required to forward the scientific understanding of energetic particles.

**Solar Energetic Particles**
The observational needs for improving the scientific understanding and forecasting of SEP events have been described in multiple white papers[1]. In general, the more of the solar surface that can be imaged and the more locations that can simultaneously sample particles and magnetic fields, the greater the opportunity to forecast and understand the physics that drives SEP events. ***A standard, versatile high-quality particle detector should be developed for use in missions of opportunity, mounted upon a crewed vehicle, or flown as a suite to locations of interest.***

---

[1] Whitman et al., "Advancing SEP Forecasting"; Whitman et al., "Sun Chaser: A Mission to the Earth-Sun Lagrangian Point 4"; Collado-Vega et al., "Space Weather Operations and the Need for Multiple Solar Vantage Points"; Raoufi et al., "Exploring the Heliosphere from the Solar Interior to the Solar Wind, Firefly".

**Magnetosphere and Van Allen Radiation Belts**

*There is a significant need to maintain consistent measurements of the radiation environment within the magnetosphere and its variability.* Following the Van Allen Probes and the soon-to-end Arase mission, radiation measurements will primarily be obtained with GOES. However, daily average fluxes derived from GOES and the Van Allen Probes can be two orders of magnitude different at similar radial distances due to strong radial gradients [77]. As solar cycle 25 increases in intensity and appears distinct from previous cycles, we thus need continued measurements throughout the magnetosphere to understand and forecast solar wind-driven magnetospheric radiation belt fluxes. A long-term mission in the radiation belts also serves as an ideal test-bed for future missions to harsher radiation environments such as Jupiter's radiation belts[2] [78]. *A follow-on mission from the Van Allen Probes has significant cross-over benefits to radiation missions in interplanetary space.*

**Spacecraft Interior and Lunar Surface**

The Gateway will be stationed in a rectilinear halo lunar orbit, ~80% of the time in interplanetary space and 20% of its time in the magnetotail [79]. This represents a major departure from crewed missions to the International Space Station. In this environment, the Gateway will sustain higher GCR intensities, which have a much larger biological effectiveness per absorbed dose compared to radiation belt particles, and will experience more extreme particle intensity and dose variations through the solar cycle.

Space weather packages will be mounted on the exterior (HERMES from NASA, ERSA from ESA) of the Gateway as well as dosimeters throughout the interior (IDA). Within materials, such as a spacecraft hull, SEPs and GCRs form secondary particles via spallation. Currently, the transport of particles through materials is understood and modeled to a limited extent [80]. Such co-located measurements are therefore important to designing future space and lunar missions. *Gateway instrumentation will provide co-located, simultaneous flux and dose measurements that should be taken advantage of to fully characterize the unique environment sampled by Gateway in cis-lunar space.*

On the lunar surface, energetic SEPs and GCRs create albedo particles resulting from interactions with the lunar regolith. These albedo particles reach energies of up to ~100 MeV, but decline rapidly in number above this energy [81]. *Previous measurements of lunar albedo neutrons only sampled up to ~20 MeV, thus future experiments should extend measurements of neutrons up to hundreds to MeV to GeV to fully characterize the lunar surface environment.*

---

[2] See also the white paper: Clark et a., "Comprehensive Observations of Magnetospheric Particle Acceleration, Sources, and Sinks (COMPASS): A Mission Concept to Jupiter's Extreme Magnetosphere to Address Fundamental Mysteries in Heliophysics"